# Multiscale modeling of heat conduction in graphene laminates

Bohayra Mortazavi[*] and Timon Rabczuk

*Institute of Structural Mechanics, Bauhaus-Universität Weimar, Marienstr. 15, D-99423 Weimar, Germany*

Abstract

We developed a combined atomistic-continuum hierarchical multiscale approach to explore the effective thermal conductivity of graphene laminates. To this aim, we first performed molecular dynamics simulations in order to study the heat conduction at atomistic level. Using the non-equilibrium molecular dynamics method, we evaluated the length dependent thermal conductivity of graphene as well as the thermal contact conductance between two individual graphene sheets. In the next step, based on the results provided by the molecular dynamics simulations, we constructed finite element models of graphene laminates to probe the effective thermal conductivity at macroscopic level. A similar methodology was also developed to study the thermal conductivity of laminates made from hexagonal boron-nitride (h-BN) films. In agreement with recent experimental observations, our multiscale modeling confirms that the flake size is the main factor that affects the thermal conductivity of graphene and h-BN laminates. Provided information by the proposed multiscale approach could be used to guide experimental studies to fabricate laminates with tunable thermal conduction properties.

*Corresponding author (Bohayra Mortazavi): bohayra.mortazavi@gmail.com

Tel: +49 176 68195567

Fax: +49 364 358 4511

# 1. Introduction

Great success of graphene [1-3] with unique combination of exceptionally high thermal [4], mechanical [5] and electrical properties [6] has raised an ongoing attention toward two-dimensional (2D) materials as a new class of materials suitable for a wide variety of applications from nanoelectronics to aerospace industry. Micro-raman spectroscopy experiments [4] confirmed that single-layer graphene can present a thermal conductivity of 4100±500 W/mK at room temperature which outperforms all known materials. Presenting a remarkably high thermal conductivity proposes the graphene as an excellent candidate in response to heat management concerns in numerous applications such as nanoelectronics and li-ion batteries. It should be taken into account that due to the small thickness of graphene sheets, application of graphene as an isolated material is complicated and limited as well. One common approach is to disperse graphene sheets inside polymeric materials to fabricate nanocomposite materials with enhanced thermal and mechanical properties [7-9]. In this regard, it is worthy to note that because of planar structure of graphene membranes, experimental tests [8] and finite element modeling [10] confirm the superiority of graphene nanosheets than carbon nanotubes in reinforcement of thermal conductivity of random nanocomposite materials.

Latest experimental study [11] however suggests a new route to reach a highly thermal conductive material through fabrication of graphene laminates. In this scenario, produced graphene flakes are closely packed and stacked to form an overlapping structure. Nevertheless, because of random nature of graphene flakes overlapping regions the physics of heat conduction in graphene laminates is non-trivial [11]. Due to difficulties of experimental studies along with complex nature of the problem, numerical and theoretical methods could be considered as promising approaches to provide a general viewpoint. Because of statistical nature of heat conduction through van-der Waals contact interfaces between individual graphene plates, application of numerical methods is more convincing and justifiable in comparison with theoretical procedures.

The objective of this work is to provide a general viewpoint concerning the heat transfer along the laminates made from 2D materials by developing a hierarchical multiscale approach. In the proposed approach, we developed a combination of atomistic and continuum methods by using molecular dynamics (MD) and finite element (FE) methods, respectively. In this regard, MD simulations were used to obtain the thermal conductivity of graphene as a function of flakes length. In addition, MD modeling was utilized to evaluate interfacial contact conductance between two separate graphene layers. Molecular dynamics simulation is a powerful tool for studying the structures response at atomistic level, however because of its high computational costs, MD turns to be an unfeasible method to study the samples with macroscopic sizes. To pass this limitation, we construct continuum models of graphene laminates by using the finite element approach. In agreement with experimental samples, we assumed that graphene flakes were randomly distributed and staked to form a representative volume element of graphene laminate. In our FE modeling, information provided by MD simulations were used to introduce flakes thermal conductivity as well as interfacial conductance between contacting sheets. It is worthy to mention that graphene is an electrically conductive material. This way, fabricated graphene laminates would be also electrically conductive due to the existence of percolation path for electrons through the contacting flakes. This issue limits the use of graphene for the applications in which the building blocks are preferred to be electrically insulating. To the best of our knowledge, hexagonal boron-nitride (h-BN) presents the highest thermal conductivity among all experimentally fabricated 2D electrically insulating materials. Consequently, the same multiscale scenario was also developed to study the heat conduction through h-BN laminates and the obtained results were compared with those of laminates made from graphene flakes.

## 2. Molecular dynamics modeling

We used classical molecular dynamics simulations in order to evaluate length dependent thermal conductivity of suspended sheets as well as thermal contact conductance between two individual flakes at room temperature (300 K). To this aim, we constructed molecular models of graphene and h-BN films. Molecular dynamics calculations in this study were performed using LAMMPS (Large-scale Atomic/Molecular Massively Parallel Simulator) [12] package. It is worthy to note that the accuracy of predicted properties by molecular dynamics simulations is strongly dependents on the use of appropriate potential functions for introducing the atomic interactions. In this work, we used optimized Tersoff potentials developed by Lindsay and Broido for graphene [13] and h-BN [14] atoms. The Tersoff potentials [15, 16] parameters proposed by Lindsay and Broido [13, 14] could predict the phonon dispersion curves of graphite and bulk h-BN in close agreements with experimental measurements. In the modeling of heat transfer between two separate layers, the van der Waals interaction are commonly introduced by Lennard-Jones (LJ) potential as expressed by:

$$\phi(r) = 4\varepsilon \left[ \left(\frac{\sigma}{r}\right)^{12} - \left(\frac{\sigma}{r}\right)^{6} \right] \qquad (1)$$

Where r refers to the distance between atoms along different sheets, ε is the depth of the potential well and σ is the equilibrium distance at which the inter-particle potential is zero. For h-BN layers, we used ε=4 meV and σ=0.3212 nm as proposed by Lindsay and Broido [14] which satisfy the measured interlayer space of 0.33 nm in bulk h-BN [17] and accurately reproduce the out-of-plane phonon dispersion in staked h-BN films. On the other hand, for the graphene flakes we applied ε=2.4 meV and σ=0.34 nm [18] which has been so far the most referred LJ potential for the modeling of carbon atoms non-bonding interactions.

In Fig.1a, the developed atomistic model for the evaluation of length dependent thermal conductivity of graphene or h-BN (Fig. 1a) films is illustrated. Fig. 2b also depicts the molecular model for the evaluation of thermal contact conductance between two separate graphene or h-BN layers. In this case, we assumed that one tenth of the sheets lengths are overlapping up-together to conduct the interfacial heat. In the molecular dynamics modeling performed in this work, we applied periodic boundary condition along the width of the samples in order to remove the effects of free atoms on the edge. These molecular models were also developed for different sheet lengths to investigate length effect on the acquired properties. We note that the time increment of all calculations was set to 1 fs.

In this work, thermal properties were evaluated using non-equilibrium molecular dynamics (NEMD) method. In this approach, we fixed atoms at the two ends of the structure to prevent atoms from sublimating (as shown in Fig. 1a with black atoms). Then, we divided the simulation box (excluding the fixed atoms at the two ends) along the longitudinal direction into several slabs. The temperature at each slab was then computed using the following relation:

$$T_i(slab) = \frac{2}{3N_i k_B} \sum_j \frac{p_j^2}{2m_j} \qquad (2)$$

Where $T_i$ (slab) is the temperature of $i^{th}$ slab, $N_i$ is the number of atoms in $i^{th}$ slab, $k_B$ is the Boltzmann's constant, $m_j$ and $p_j$ are atomic mass and momentum of atom j, respectively. In the NEMD method, specimen was first relaxed at room temperature using Nosé-Hoover thermostat (NVT) method for 100 ps. In the next step, a temperature difference was applied between the first and last slabs using the NVT method, while the remaining slabs were under constant energy (NVE) simulations. In this study, the first and last slabs were assigned to be hot and cold reservoir, respectively (as shown in Fig. 1). To keep the applied temperature differences at the two slabs at the ends, at every simulation time step an amount of energy is added to the atoms in hot reservoir and at the same time another amount of energy is removed from the atoms in the cold reservoir by the NVT method. Instantly after the application of temperature difference, the system would be at transient condition. However, after 300 ps of the exchanging process, the system reaches to a steady-state heat transfer condition in which a temperature gradient is established along the sample longitudinal direction. At this stage, simulations were performed for longer time and averaged temperatures at each slab were computed (as shown in Fig. 2a and Fig. 2b). To ensure the accurate application of NEMD method after reaching the steady-state heat transfer condition, the net amount of energy added to the hot reservoir must be equal to that removed from the cold reservoir. This means that the total energy of the system is conserved. In addition, the slopes of energy curves must follow linear patterns to ensure that the energy is added or removed with a constant rate. In Fig. 2c and Fig. 2d, we plot the energy curves for the evaluation of length dependent thermal conductivity of pristine sheets as well as those for studying the thermal contact conductance, respectively. As it can be observed, in both cases the slopes of energy curves are considerably close and they follow linear patterns. These observations imply that the energy of the simulation box remains constant and a constant heat flow is applied throughout the sample (Fig. 1).The applied heat flux through the sample ($q_x$) is calculated based on the slope of these energy curves which is accurately the same if one calculates it from the hot or cold atoms energy curves.

Based on the established temperature profiles along the samples and calculated heat fluxes, we then evaluated the thermal conductivity and contact conductance. The thermal conductivity, *k*, was calculated using the one-dimensional form of the Fourier law:

$$q_x = -kA_{cross}\frac{dT}{dx} \qquad (3)$$

Here, $\frac{dT}{dx}$ is the established temperature gradient along the sample (as shown in Fig. 2a with dashed line) and *A$_{cross}$* is the cross sectional area of graphene or h-BN flakes. In the calculation of cross section area, we assumed thickness of 0.34 nm [18] and 0.33 nm [17] for graphene and h-BN sheets, respectively. On another hand, thermal contact conductance, *C*, was also evaluated using the following relation:

$$q_x = CA_{contact}\Delta T \qquad (4)$$

Here, ΔT is the formed temperature difference at the interface of two membranes (as depicted in Fig. 2b) and $A_{contact}$ is the overlapping area of two contacting films.

## 3. Finite element modeling

In the multiscale scheme developed in this work, we used finite element method to evaluate the effective thermal conductivity of graphene and h-BN laminates. We note that computational limitation of finite element method impose limits on the size of studied systems. Therefore, commonly the simulations are limited only to the modeling of a representative volume element (RVEs), consisting of a limited number of particles. Finite element modeling in this study was performed using the Abaqus/Standard (Version 6.10) commercial package. As a common assumption, graphene and h-BN sheets were modeled using the disc geometry. In this case, the diameter of the disc is considered as the flakes length. Fig. 3a illustrates a sample of developed RVE of graphene laminate that is consisting of 800 individual flakes staked in 12 layers up together. In agreement with experimental samples, the flakes were randomly distributed in each layer. The constructed models were periodic in loading direction (y direction in this case). These models were constructed in Abaqus by python scripting. We remind that the acquired results by MD modeling were used to introduce flakes thermal conductivity as well as contact conductance interaction properties. For the evaluation of laminates effective thermal conductivity, we included two highly conductive strips at the two ends of the structure. These two auxiliary parts had the same cross section as that of the laminate and they were thermally tied (zero contact resistance) to the contacting flakes at the two ends. The thermal conductivity of these thin strips was chosen to be one million to have negligible resistance. For the loading conditions, a constant inward surface heat flux (+$q_y$) was applied on the external surface of a strip (hot surface) while on the outer surface of the opposite strip (cold surface) the same magnitude outward surface heat flux (-$q_y$) was applied. As the initial value for the problem, we set the temperature of outer surface of cold strip to zero. As a result of applied loading condition, a steady-state temperature profile form along the RVE length (Fig. 3b). We note that heat percolates in laminate only through the contacting surfaces. To this purpose, we introduced contact elements between every two contacting sheets. In this case, the master surface was chosen to be the one that is closer to the hot strip and then the other contacting surface (closer to the cold strip) was selected to be the slave surface. The established temperature difference along the laminate, $\Delta T_l$, is therefore the computed temperature on the outer surface of hot strip. Finally, the laminate effective thermal conductivity, $k_{eff}$, was obtained using one-dimensional form of the Fourier law:

$$k_{eff} = q_y \frac{L_l}{\Delta T_l} \qquad (5)$$

Here, $q_y$ is the applied surface heat flux and $L_l$ is the laminate length (without the strips).

## 4. Results and discussions

In Fig. 4, the calculated thermal contact conductance for graphene and h-BN flakes are plotted for different sheets lengths. As it can be observed, the obtained results are independent of flake length. On the other hand, as we assumed that one tenth of sheets lengths are overlapping, the contacting area increases by increasing the sheet length. Accordingly, the acquired contact conductance values are also independent of contacting area. To summarize, using the NEMD method we found the conduct conductance of 0.1±0.01 GW/m$^2$K and 0.053±0.005 GW/m$^2$K for h-BN and graphene flakes, respectively. These values were then used in our finite element modeling to introduce contact interaction properties between the flakes, independent of flakes length. We note that among the different methods available for the modeling of interfacial contact conductance, the NEMD approach is the most convincing approach that could simulate the nature of heat conduction in graphene and h-BN laminates.

We performed NEMD simulations for samples with different lengths up to 300 nm. As illustrated in Fig. 5, MD results for graphene and h-BN sheets show that the thermal conductivity increases by increasing the length (L). As a common approach, the thermal conductivity of sheets with infinite length was obtained by extrapolation of results for finite lengths. To this aim, a simple approach is to define an effective phonon mean free path ($\Lambda_{eff}$) as $1/\Lambda_{eff} = 1/\Lambda + 1/L$. Since the thermal conductivity is proportional to $\Lambda_{eff}$, the thermal conductivity of the infinite system can be obtained by extrapolating to $1/L=0$ [20]. Accordingly, the thermal conductivity of graphene and h-BN at room temperature are calculated to be 3050±100 W/mK and 600±20 W/mK, respectively. The NEMD result for thermal conductivity of single-layer h-BN at room temperature is in an excellent agreement with theoretical prediction of 600 W/mK by Lindsay and Broido [14]. We note that using the Tersoff potential parameters proposed by Matsunaga *et al.* [21], thermal conductivity of pristine h-BN was predicted to be 80 W/mK [22]. On the other hand, the calculated value for graphene is below the experimental measurements of 4100±500 W/mK [4] for pristine and single-later graphene. The obtained relations for thermal conductivity as a function of flakes length were then used to introduce the intrinsic thermal conductivity of particles in our finite element modeling.

In Fig. 6, we compared the multiscale results for effective thermal conductivity of graphene and h-BN laminates for different flake sizes. We note that we constructed finite element models for different number of layers and we found that when the numbers of laminate layers are higher than eight, the predicted properties are well-converged. However, in the final calculation in this work, we set the laminate layers to twelve. Our multiscale results confirm that by increasing the flake size, the effective conductivity increases for both graphene and h-BN laminates. This observation is in agreement with recent experimental results [11] in which it was observed that the flakes size presents remarkable effects on thermal conductivity of graphene laminates. This enhancement trend in thermal conductivity is not only because of the increasing of flakes thermal conductivity for the larger flakes but it is strongly due the increasing of contacting area for heat percolation. In Fig. 6, the multiscale results are also compared with experimental measurements for graphene laminates [11]. Interestingly, the averaged thermal conductivity of three experimental points is only within 2% difference with the multiscale prediction for the same averaged flake size. This confirms remarkable validity of the developed hierarchical multiscale method in this study. However, experimental results show sharper effects on the laminate thermal conductivity as the flake size increases. In the reported experiments only about 20% increase in the flake size resulted in a two times higher thermal conductivity of graphene laminate. This considerable increase could not be due to the changes in the intrinsic thermal conductivity of the flakes [23-25]. However, one must take into account that in experiments it is not possible to change the average flake size while keeping all other parameters of the composite exactly the same. This means that variations in the flake overlap or the strength of their attachment, which define the interfacial conductance between the flakes, can account for some discrepancy between the experimental and numerical results.

Surprisingly, our multiscale results suggest that for the flakes sizes lower than 1μm, the h-BN laminates present higher thermal conductivity than graphene laminates. This observation is interesting since the intrinsic thermal conductivities of h-BN films are by several times lower than that for graphene sheets (as compared in Fig. 5). We remind that our MD results show that interfacial contact conductance between h-BN films are almost double of that for graphene sheets. Therefore, superior thermal conductivity of h-BN laminates for small flake sizes is principally due to the lower interfacial resistance between the h-BN particles. In this case, for the small flake sizes the interfacial resistance is the main factor that does contribute to the effective thermal conductivity of laminate and the particles thermal conductivity plays the minor part. However, by increasing the flake size, effect of interfacial resistance is weakened and on the other hand influence of particles intrinsic conductivity is enhanced. Our results suggest that a laminate made by graphene flakes with size of 10 μm could yield a thermal conductivity by around 2.5 times higher than a laminate formed from same sized h-BN films. In addition, our multiscale results predict that a graphene laminate formed by millimeter sized graphene sheets can exhibit a remarkably high thermal conductivity of around 900 W/mK. This thermal conductivity is by around thirteen times higher than that of the sample with the flake size of 1μm. Such an observation clearly shows the remarkable importance of flake size effects. We note that the intrinsic thermal conductivity of flakes with length of 1μm is only 25% below the one with 1 mm length. On the other hand, h-BN films could be also considered as an excellent candidate to fabricate highly conductive laminates. Owing to their low interfacial thermal resistance, h-BN laminates are best suited when the flake sizes are small.

From experimental point of view, fabrication of graphene or h-BN films with large sizes has been among the most challenging barriers front to their applications performance. Nevertheless, our simulation results recommend that for the small flakes sizes, improvement of interfacial conductance could be taken into consideration as a promising approach to enhance the effective thermal conductivity of laminates. In this regard, the common route is to form covalent bonds between the graphene flakes through chemical functionalization or ion irradiation. However, it should be taken into account that these chemical modifications also decrease the intrinsic thermal conductivity of graphene sheets. Our recent MD study [10] predicts that formation of 5% covalent bonds between graphene and epoxy enhances the thermal contact conductance by around 3 times and reduce the intrinsic thermal conductivity of pristine films by around 60%. As an example and based on these estimates, our multiscale modeling reveals that a graphene laminate with flake size of 1μm and incorporation of 5% covalent bonds could yield a thermal conductivity by around 70% higher than the same sample made only from pristine sheets. Because of the fact that contact resistance between the particles is understood to be the main factor in determining the overall effective thermal conductivity for smaller flake sizes, effect of interlayer covalent bonding intensifies more as the film sizes are smaller.

Presented multiscale results suggest the possibility of tuning the thermal conductivity of graphene laminates by almost two orders of magnitude through controlling the flakes size. Therefore, studying the electrical conductivity of graphene laminates could be also interesting. This is due to the fact that fabrication of graphene laminates with low thermal and high electrical conductivities could enhance the thermoelectric figure of merit in favor to reach a high efficiency carbon-based thermoelectric material.

## 5. Conclusion

First multiscale modeling was conducted to systematically study the effective thermal conductivity of graphene and h-BN laminates. The proposed approach is a hierarchical combination of atomistic and continuum modeling. We performed molecular dynamics simulations for the evaluation of length dependent thermal conductivity of particles and thermal contact conductance between two individual flakes. Then, we constructed continuum models of graphene and h-BN laminates using the finite element approach. In our finite element modeling, we used the information provided by atomistic simulations to introduce particles as well as interfacial conductance thermal properties. Comparison with experimental results for graphene laminates confirms the considerable validity of the proposed approach.

Our modeling results suggest the possibility of tuning the graphene laminates thermal conductivity by almost two orders of magnitude through controlling the flakes size. We found that for the small flake sizes, the inter-flake contact resistance presents the major contribution toward the effective thermal resistance of the sample. Unexpectedly, we found that for the flake sizes smaller than 1μm, h-BN laminates owing to their high interfacial conductance could yield thermal conductivity higher than the laminates made form graphene sheets. Our study recommends that forming the covalent bonds between the graphene flakes could be considered as an efficient approach to considerably enhance the thermal conductivity of laminates, especially when the flake size is small. Nevertheless, by increasing the flake size, the intrinsic thermal conductivity of particles start to play the key role in the heat conduction, in which we found superiority of graphene than h-BN laminates for the flake sizes larger than 1μm. The proposed multiscale approach in this study could be considered as an efficient modeling methodology to design laminates with tunable thermal conduction properties.

**Acknowledgment**

Financial support by European Union through ERC grant for COMBAT project is greatly acknowledged.

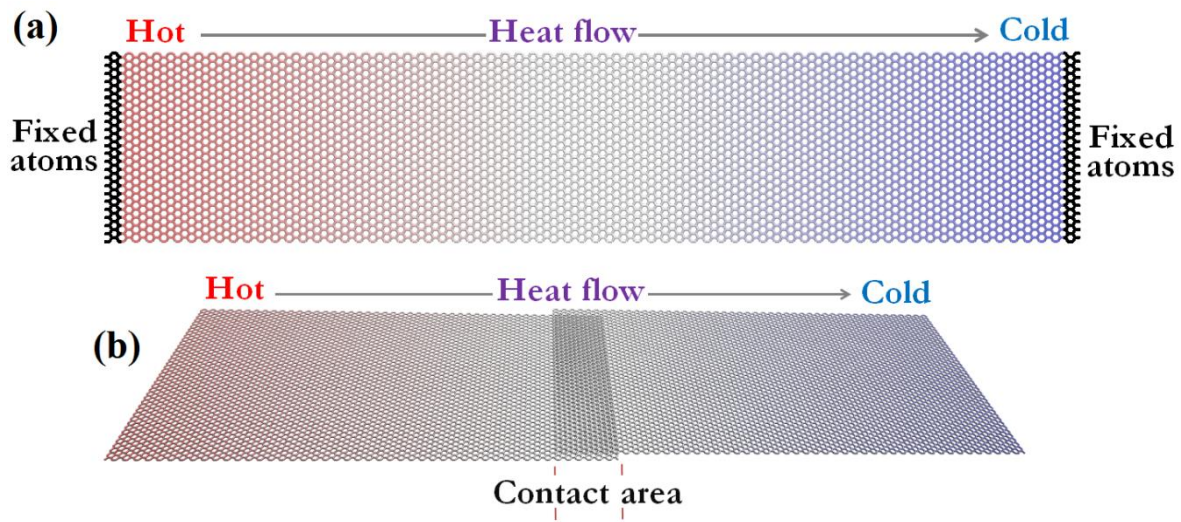

Fig. 1- Constructed molecular models for the valuation of (a) length dependent thermal conductivity and (b) thermal contact conductance between two individual flakes. Modelings were performed for defect-free graphene and h-BN films. For the modeling of thermal contact conductance, one tenth of sheets lengths are overlapped to simulate interfacial heat conductance. VMD software [19] is used for graphical presentation of structures.

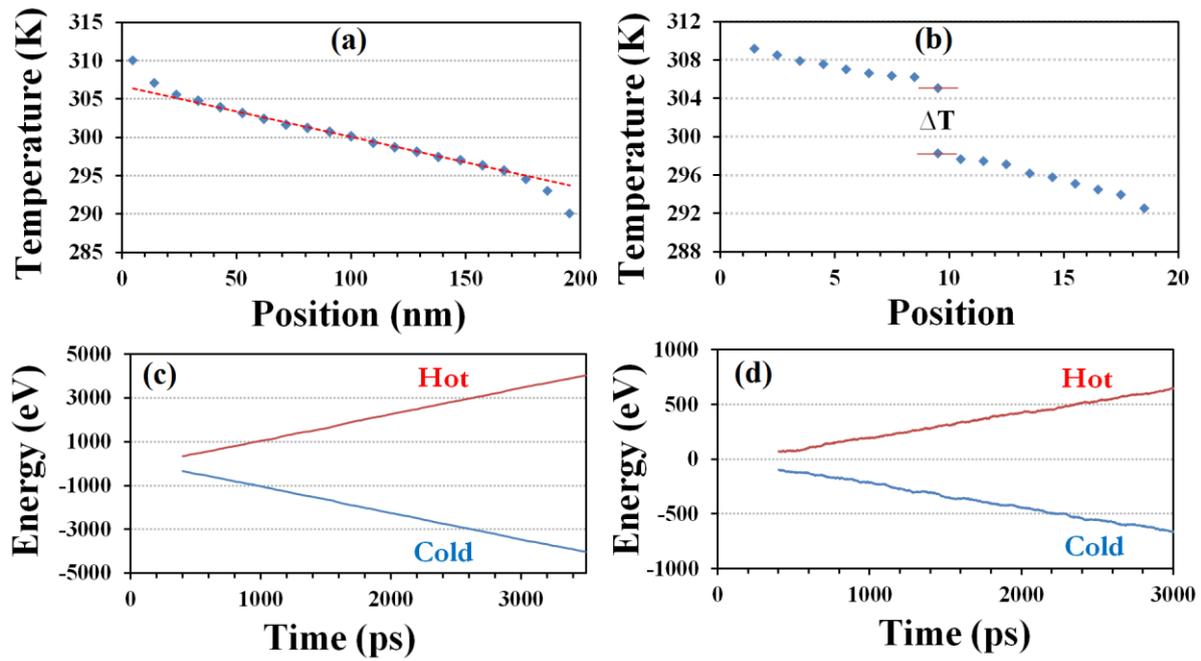

Fig. 2- (a and b) Established steady-state temperature profile along the samples by application of non-equilibrium molecular dynamics method. (a) In the modeling of length dependent thermal conductivity, by neglecting the initial jumps at two ends, a linear relation (dT/dx) is formed along the sample length that is shown by a dashed line. (b) For the modeling of thermal contact conductance, a temperature jump (ΔT) occurs at the interface of two individual flakes. Samples of energy values added to the atoms in hot reservoir and removed from the atoms in cold reservoir by the NVT method for the modeling of (c) length dependent thermal conductivity and (d) contact conductance. The depicted results (a and c) belong for the thermal conductivity modeling of a graphene sheet with the length of 200 nm and (b and d) are for the modeling of contact conductance between two individual h-BN films, each with the length of 100 nm.

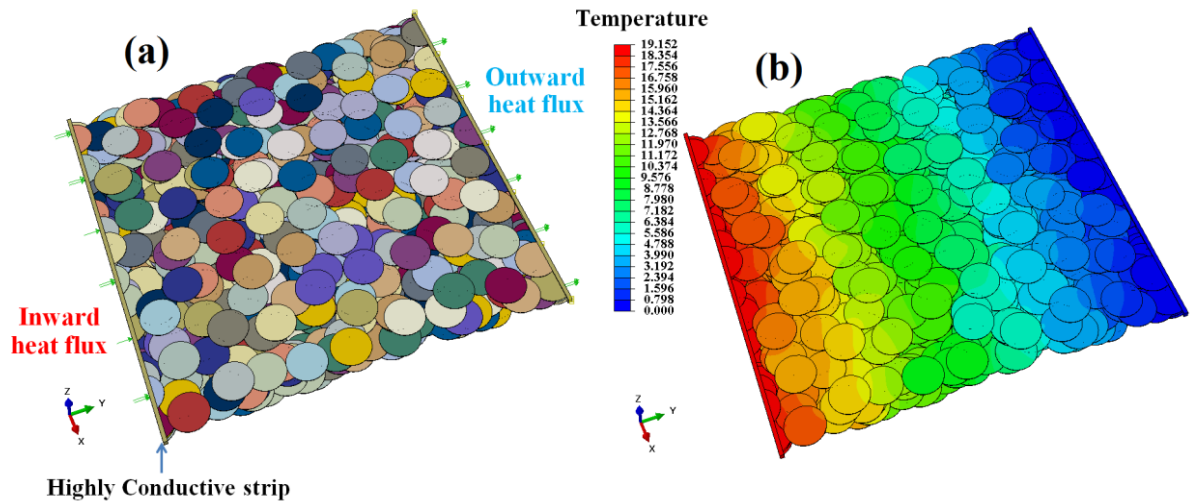

Fig. 3- A typical developed 3D representative volume element of graphene laminate constructed in Abaqus/Standard. (a) For the loading condition, we applied inward and outward surface heat fluxes on the outer surfaces of highly conductive strips. (b) Established steady-state temperature profile along the sample.

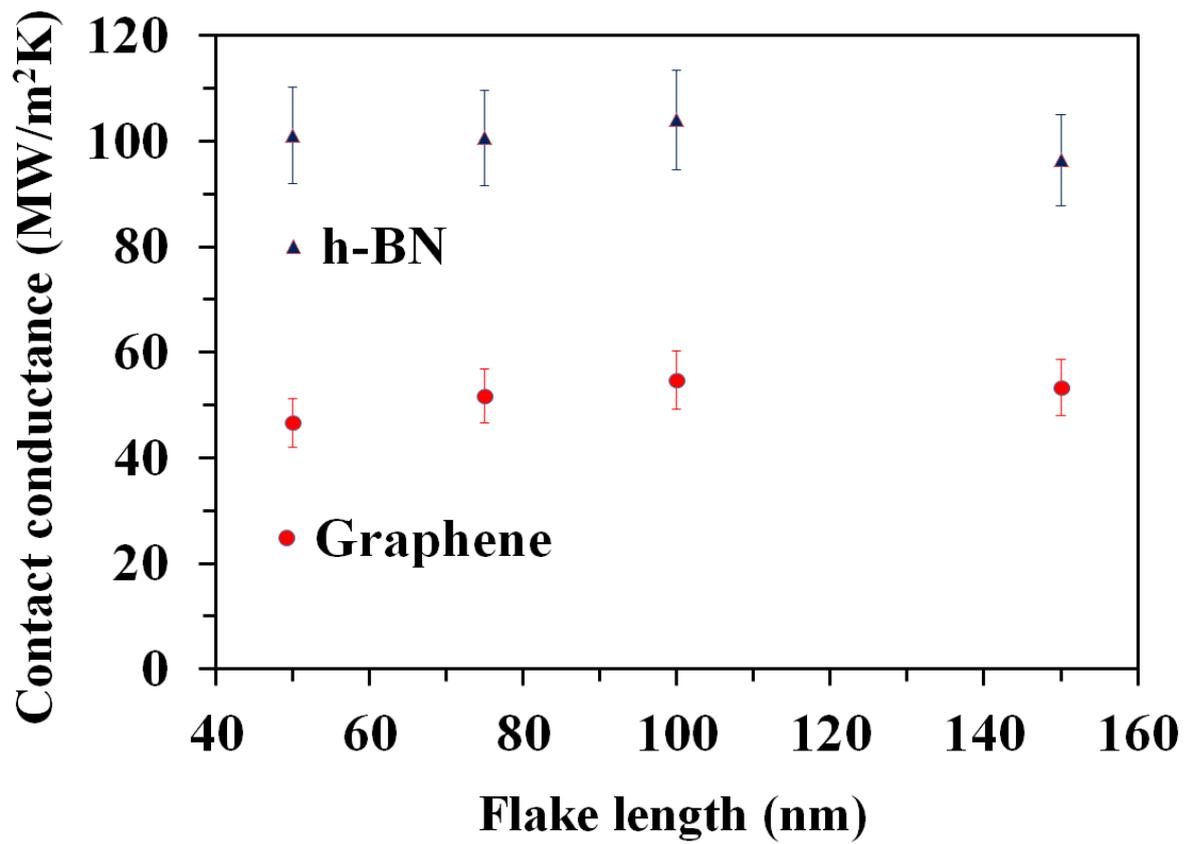

Fig. 4- Calculated thermal contact conductance between individual graphene or h-BN sheets for different sheet lengths.

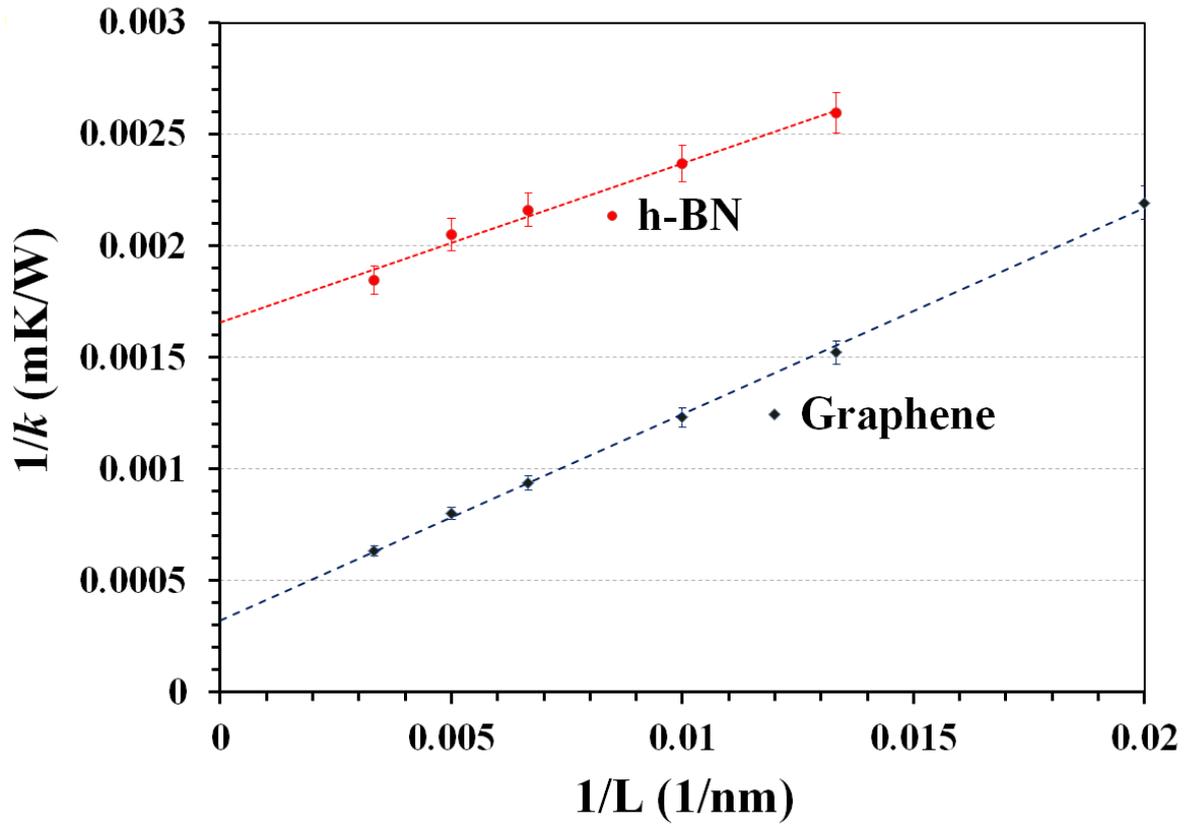

Fig. 5- Calculated length dependent thermal conductivity of pristine graphene and h-BN sheets. We plot the inverse of thermal conductivity as a function of length inverse to extrapolate the thermal conductivity of films with infinite lengths. The fitted lines to the NEMD results (shown by dashed lines) are also used to determine length dependent thermal conductivity functions for graphene and h-BN films.

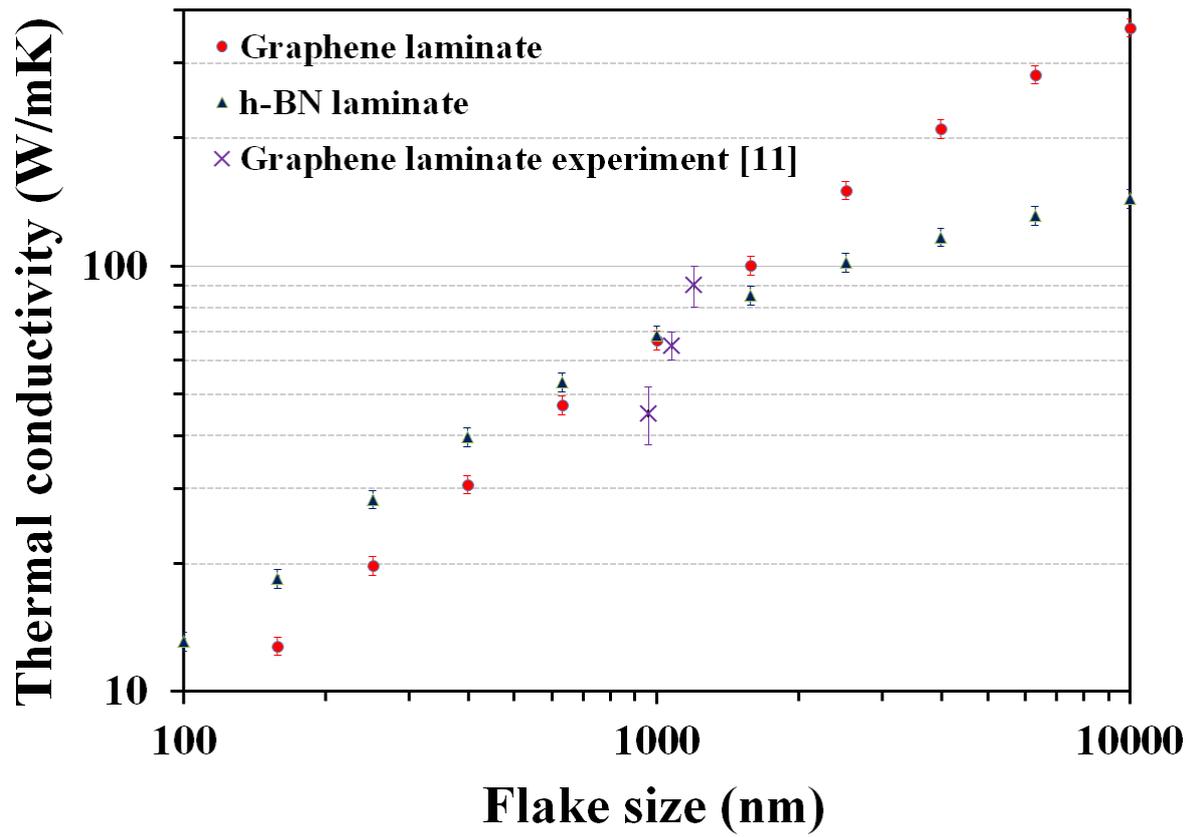

Fig. 6- Multiscale results for effective thermal conductivity of graphene and h-BN laminates as a function of flake size. Modeling results are compared with recent experimental measurements [11] for graphene laminates.